\documentclass[preprint,12pt]{elsarticle}   
\usepackage{amssymb}
\usepackage{xurl}
\usepackage{comment}
\usepackage{multirow}
\usepackage{tabularx}
\usepackage{amsmath}
\usepackage{enumitem}
\usepackage{hyperref}
\usepackage{bookmark}
\usepackage{xcolor}
\usepackage{wasysym}
\usepackage{graphicx}
\usepackage{subfig}
\usepackage{multirow}
\usepackage{svg}
\usepackage{pdfpages}

\journal{}

\begin{document}

\begin{frontmatter}
\title{A survey on factors influencing mobile application usability through the lens of PACMAD+3 model}
%\begin{comment}
\author[inst1]{Paweł Weichbroth}
\affiliation[inst1]{organization={Gdansk University of Technology, Faculty of Electronics, Telecommunications and Informatics, Department of Software Engineering},
            city={Gdansk},
            country={Poland}
            email: pawel.weichbroth@pg.edu.pl}
%\ead[inst1]{pawel.weichbroth@pg.edu.pl}
%\end{comment}

\begin{abstract}
Undeniably, the advent of mobile applications has brought new frontiers to usability engineering. To date, ongoing research has shown significant efforts to adopt and adapt usability principles to the mobile computing environment. One of these endeavors is the PACMAD+3 model. However, to the best of our knowledge, little or no effort has been made to empirically evaluate these factors against perceived influence. With this in mind, the objective of this study is to explore this issue. To achieve this goal in a reliable and reproducible manner, we took advantage of previous attempts to conceptualize the mobile usability factors, but we contribute by operationalizing these theoretical constructs into observable and measurable phenomena. In this sense, the survey was designed and carried out on a sample of 838 users to assess the significance of the PACMAD+3 factors on the perceived usability of mobile applications. Our findings show that, on average, users rated efficiency as highly important, while the remaining seven, namely: cognitive load, errors, learnability, operability, effectiveness, memorability, and understandability, were rated moderately important. Insights into the importance of usability factors and the corresponding features can also facilitate the design and development of mobile applications. Therefore, our research contributes to the field of human-computer interaction with theoretical and practical implications for mobile usability researchers, UX designers, and quality assurance engineers.
\end{abstract}


\end{frontmatter}

\section{Introduction}
\label{sec:introduction}
The introduction of the first smartphone in 2007 marked has changed the world computed and communicated \cite{ens2012guidelines}. It is well known that the "reinvention of the phone" has led to a combination of hardware and software features that have taken the quality of use to new heights \cite{moumane2016usability}. The benefits of smartphones include instant communication, digital consumption of social media, gaming and entertainment, just to name a few.

From the perspective of software vendors, the mobility paradigm introduces a different environment for applications, requiring significant changes in their design and development \cite{lee2004mobile, wasserman2010software}. Thus, usability has been rethought in the light of limited resources on the one hand and different user requirements on the other. However, little is known about the factors and specific features that influence the perceived usability of mobile applications. 

While for user experience practitioners the term "usability" as a goal is equivalent to quality of use, which means that the product is used by real and satisfied users \cite{nascimento2016userbility, jeddi2020usability}. 
In this way of thinking, usability has two equivalent roles in design: as an attribute that must be designed into a product, and as the highest quality goal that ultimately aims at user satisfaction. For users of mobile applications, usability goes beyond mere functionality and includes factors related to a seamless and enjoyable interaction with the application \cite{adu2020development, manzano2023usability, shinozaki2024usability}. On the other hand, poor usability can lead to frustration \cite{raita2014mixed, sarkar2016usability, cho2018multi}, lower user engagement \cite{huang2023systematic, weichbroth2025usability}, and ultimately abandonment \cite{liang2024exploratory, al2023people} in favor of competitors.

There have been a number of intensive research efforts to understand the nature and phenomena of mobile usability \cite{aryana2013mobile, da2019set, kivijarvi2023instrumental}, with a particular focus on identifying contributing factors \cite{ham2006conceptual, baharuddin2013usability, ghazizadeh2017quantitative}. 
However, despite nearly two dozen years of studies, the ongoing discussion has not yielded conclusive results due to the lack of uniformity in modeling mobile usability. One of the first attempts to find a unified model was made in 2013 by Harrison et al. \cite{harrison2013usability}, who introduced the PACMAD model. Due to its structured approach to identifying and addressing usability issues specific to mobile applications, their model has become one of the most widely used frameworks for mobile usability testing \cite{rey2024ligtas, weichbroth2024usability}.

In 2024, PACMAD was updated by Weichbroth \cite{weichbroth2024usability} to PACMAD+3, incorporating three additional factors to reflect recent technological advances in both hardware and software area. However, to the best of our knowledge, no study has yet  attempted to evaluate the impact of all these factors on the perceived usability of mobile applications. 
In order to fill this gap and provide a deeper understanding in this area, our study is guided by this concern and seeks to answer it through a literature review and survey. In short, while the review aims to conceptualize and operationalize both usability factors, the questionnaire aims to evaluate a set of features by users of mobile applications. In addition, the novelty of this paper lies in the operationalization of the usability factors in terms of observable and tangible qualities or specific user interface components. 

\section{Background and Motivation}
\label{Background}
In light of the recently published results  \cite{weichbroth2020usability}, the most commonly adopted usability definition for mobile applications is ISO 9241-11, which is also considered valid in the current study.
Hence, usability is understood as the "extent to which a system, product or service can be used by specified users to achieve specified goals with effectiveness, efficiency and satisfaction in a specified context of use" \cite{ISO9242-11}.

For the sake of clarity, while the majority of studies refer to usability attributes as the notion of certain qualities of mobile applications, we will refer to usability factors in this paper due to the exclusive focus on user perceptions. Each factor is then manifested by a set of mobile application features or properties. 

According to ISO 9241-11 \cite{ISO9242-11}, the concept of usability can be broken down into three factors, namely:
\begin{itemize}
    \item effectiveness: “the accuracy and completeness with which users achieve specified goals”;
    \item efficiency: “the resources expended in relation to the accuracy and completeness with which users achieve goals”;
    \item satisfaction: “the comfort and acceptability of the work system to its users and other people affected by its use”.
\end{itemize}

Considering the above definitions from the user's perspective, only effectiveness and efficiency are subject to evaluation due to their nature. In other words, satisfaction is excluded from our study as it is a result of the physical interaction between a user and a mobile application.

It should be noted that the "context of use" is defined as the "characteristics of users, tasks, and organizational and physical environments" \cite{ISO9242-11}. It is a critical concept in human-centered design and mobile usability studies. It refers to the specific environment, conditions, and characteristics in which a product, system, or interface is used. Understanding the context of use is essential for usability design and evaluation to be relevant and actionable.

The literature review shows the lack of consensus regarding the usability attributes of mobile applications. It should be noted that a lot of research has been carried out in this area. Researchers have defined usability itself in different ways and specified different attributes. However, the recent study conducted by Weichbroth \cite{weichbroth2024usability} has informed our research as this paper attempts to find a consensus by synthesizing the state of the art literature in the field of mobile usability. 
In particular, the paper introduces a consolidated universal usability model for mobile applications, termed PACMAD+3. Through the lens of existing human-computer interaction theory, and by employing a mix of qualitative and quantitative methods, this model appears to synthesize the current body of knowledge in a comprehensive and reliable manner. 

The PACMAD+3 model consists of ten attributes divided into three groups:
\begin{itemize}
    \item ISO 9241-11: effectiveness, efficiency, satisfaction.
    \item PACMAD: learnability, memorability, errors, cognitive load.
    \item Others: understandability, ease of use, operability.
\end{itemize}

In summary, Table~\ref{tab:SOTA-models} presents the state-of-the-art usability models and attributes for mobile applications.

\begin{table}[h]
\caption{The state-of-the-art usability models and attributes for mobile applications.}
\label{tab:SOTA-models}
\centering
\small
\begin{tabular}{|l|llc|}
\hline
\multirow{2}{*}{\textbf{Attribute}} & \multicolumn{3}{c|}{\textbf{Model}}                                                 \\ \cline{2-4} 
   & \multicolumn{1}{l|}{ISO 9241--11} & \multicolumn{1}{l|}{PACMAD} & PACMAD+3 \\ \hline
effectiveness              & \multicolumn{1}{c|}{$\bullet$}       & \multicolumn{1}{c|}{$\bullet$} & $\bullet$   \\ \hline
efficiency                 & \multicolumn{1}{c|}{$\bullet$}       & \multicolumn{1}{c|}{$\bullet$} & $\bullet$   \\ \hline
satisfaction               & \multicolumn{1}{c|}{$\bullet$}       & \multicolumn{1}{c|}{$\bullet$} & $\bullet$   \\ \hline
learnability               & \multicolumn{1}{c|}{}             & \multicolumn{1}{c|}{$\bullet$} & $\bullet$   \\ \hline
memorability               & \multicolumn{1}{l|}{}             & \multicolumn{1}{c|}{$\bullet$} & $\bullet$   \\ \hline
errors                     & \multicolumn{1}{l|}{}             & \multicolumn{1}{c|}{$\bullet$} & $\bullet$   \\ \hline
cognitive load             & \multicolumn{1}{l|}{}             & \multicolumn{1}{c|}{$\bullet$} & $\bullet$   \\ \hline
understandability          & \multicolumn{1}{l|}{}             & \multicolumn{1}{c|}{}       & $\bullet$   \\ \hline
ease of use                & \multicolumn{1}{l|}{}             & \multicolumn{1}{c|}{}       & $\bullet$   \\ \hline
operability                & \multicolumn{1}{l|}{}             & \multicolumn{1}{c|}{}       & $\bullet$   \\ \hline
\end{tabular}
\end{table}

Given the context of the current research, we also exclude ease of use for similar reasons in the case of satisfaction. Therefore, a total of eight attributes are further conceptualized and operationalized. 
For the sake of clarity, in the rest of the paper we will refer to an attribute as a factor.

\section{Factors Conceptualization and Operationalization}
This section is divided into subsections that reflect unique usability factors. First, a factor definition is provided along with commonly used measurement metrics. Second, each factor is then operationalized by a set of manifest variables that correspond to a particular feature of a mobile application or the result of an interaction between a user and a mobile application. Each particular variable is assigned a unique three-character code, where the first two characters are the short name of the factor (fixed part) and the last character is the number (dynamic part). To present this information in a readable way, we used a tabular format, easy to interpret and less time consuming.

\subsection{Effectiveness}
Effectiveness is the ability of a user to complete a task in a specified context \cite{harrison2013usability, parsazadeh2018framework}, and is measured by assessing whether or not participants can complete a set of specified tasks to estimate the task completion rate \cite{miguel2017voice}. 

From the user perspective, the ability to perform a given task is supported by four application properties, given in Table \ref{tab:oper-effectiveness}.

\begin{table}[h]
\caption{Mobile Effectiveness Operationalization.}
\label{tab:oper-effectiveness}
\footnotesize
\begin{tabular}{|l|p{9cm}|p{2.5cm}|}
\hline
\textbf{Code}  & \textbf{Feature}     & \textbf{Source} \\ \hline
ES1 &  total number of steps required to complete the task & \cite{restuputri2022role}, \cite{weichbroth2020usability} \cite{weichbroth2022empirical}        \\ \hline
ES2 &  autofill form feature      &  \cite{rukzio2008automatic}, \cite{oesch2021emperor}, \cite{simmons2021systematization}      \\ \hline
ES3 &  automatic login   &   \cite{zhu2019riskcog}, \cite{shin2023heuristic}      \\ \hline
ES4 &  access to frequently used functions & \cite{mahamad2007user}, \cite{kern2009design}, \cite{yang2022influence} \\ \hline
\end{tabular}
\end{table}

In mobile devices, a task is considered to be efficiently designed if a number of required actions is as low as possible for a user to perform. Having said that the following three mobile application features facilitate task performance.  
First, the autofill feature of the mobile application has been introduced to save time and maximise convenience, during the task performance by the user. Second, with auto-login enabled, an application will pop-up the user's credentials whenever the user visits a login page, saving time and effort by eliminating the need to enter a username and password. 
Third, the three-bar ("hamburger") menu navigation, which facilitates access to frequently used functions, has become widespread due to the physical limitations of small smartphone screens.

\subsection{Efficiency}
Efficiency is the ability of the user to complete their task with speed and accuracy \cite{nosseir2012mobile, hutahaean2020identifying}. This attribute reflects the productivity of a user while using the application \cite{bevan2008classifying}. Efficiency can be measured in a number of ways, such as the time to complete a given task, or the number of keystrokes required to complete a given task. These requirements are supported by four application properties, given in Table \ref{tab:oper-efficiency}.

\begin{table}[h]
\caption{Mobile Efficiency Operationalization.}
\label{tab:oper-efficiency}
\footnotesize
\begin{tabular}{|l|p{9cm}|p{2.5cm}|}
\hline
\textbf{Code}  & \textbf{Feature}     & \textbf{Related Work} \\ \hline
EY1 & duration of the application starting and closing & \cite{dixon2016usability}, \cite{nam2016user}, \cite{weichbroth2024usabilitya}        \\ \hline
EY2 & duration of content loading   & \cite{barakovic2015multidimensional},   \cite{wang2021effect}, \cite{wang2023waiting}  \\ \hline
EY3 & application performance continuity (no deadlocks) &  \cite{gao2014mobile}, \cite{majrashi2018task}      \\ \hline
EY4 & duration of the application response to the performed action & \cite{von2016usability}, \cite{saleh2017evaluating}, \cite{yu2020unravelling}  \\ \hline
\end{tabular}
\end{table}

Duration is the primary metric for evaluating the performance of a mobile application. In this context, the measurement refers to both the launch and shutdown time of the mobile application. Besides, depending on the its nature, other areas may also be monitored and acted upon to optimize performance by mitigating resource consumption. 
A deadlock in the mobile computing environment refers to the state where two or more threads are incapable of proceeding because each is waiting for the other to release an occupied resources, preventing any progress \cite{Hussain2024}. One of the most common effects of deadlocks is application performance degradation, manifested by significant response delays that can affect other running applications. Other more severe effects include application crashes, hangs, unresponsiveness or, in the worst case, fatal errors. 

\subsection{Learnability}
Learnability is the ease with which a user can gain proficiency with an application \cite{az2019evaluating}. 
It typically reflects how long it takes a person to be able to use the application effectively \cite{irwansyah2018radio}. In order to measure learnability, a user’s performance during a series of tasks is observed to measure how long it takes these participants to reach a pre-specified level of proficiency. 

Table \ref{tab:oper-learnability} presents app features support learning user's capabilities.
\begin{table}[h]
\caption{Mobile Learnability Operationalization}
\label{tab:oper-learnability}
\footnotesize
\begin{tabular}{|l|p{9cm}|p{2.5cm}|}
\hline
\textbf{Code} & \textbf{Feature}    & \textbf{Related Work} \\ \hline
LY1 & intuitive use  &  \cite{marsh2008design}, \cite{britton2013intuitive}, \cite{naumann2010benchmarks}      \\ \hline
LY2 & learning duration of how to use the application functions & \cite{liu2021case},  \cite{ramdowar2023comprehensive}, \cite{xiao2024design} \\ \hline
LY3 & changes resulting from the application update &  \cite{mathur2017impact}, \cite{huang2022updating}      \\ \hline
\end{tabular}
\end{table}

Under the umbrella of mobile usability, the key principle revolves around learnability, emphasizing the ease with which visitors can accomplish tasks during their first interaction. Essentially, the mobile application should be effortless to understand and navigate, achieved through intuitive design, instinctive navigation, and informative prompts to help users navigate the interface. 
It is well known that reducing learning time increases perceived usability by raising task completion effectiveness and ultimately user satisfaction.

\subsection{Memorability}
Memorability is the degree to which a user can remember how to use an application in an effective way \cite{weichbroth2020usability} which is the result of the ability to recognize its functions and menus \cite{nurdina2021usability}.
Generally, some applications are used sporadically while some are used on a regular basis. However, both are expected to be used without the need to learn and remember how to use them. Human perception is developed by a combination of attention, eye movements and memory \cite{lahrache2018visualizations}. Thus, memorability can be measured by asking users to perform a series of tasks after having become proficient with the use of the application, and afterwards asking them to perform similar tasks after a period time of inactivity. To determine how memorable the application was a comparison can be made between the two sets of results \cite{harrison2013usability}. 

Table \ref{tab:oper-memo} shows an app properties which supports user's ability to memorize its effective usage.

\begin{table}[h]
\caption{Operationalization of the attribute Memorability for mobile applications}
\label{tab:oper-memo}
\footnotesize
\begin{tabular}{|l|p{9cm}|p{2.5cm}|}
\hline
\textbf{Code}  & \textbf{Feature}   & \textbf{Related Work} \\ \hline
MY1 & using the application does not require memorizing its specific options, messages and symbols & \cite{coursaris2011meta}, \cite{afif2021evaluating}, \cite{nizamani2021novel}      
\\ \hline
MY2 & using the application does not require memorizing previously input data &  \cite{abd2019evaluation}, \cite{hsu2014usability}      
\\ \hline
MY3 & displaying task hints &   \cite{hung2012enhancing}, \cite{bodrunova2018impact}     \\ \hline
MY4 & description (metadata) of locally stored data & \cite{nickerson2015managing}, \cite{nickerson2016selecting}, \cite{azadi2020mobile}  \\ \hline
\end{tabular}
\end{table}

By setting a common language and providing standard reference points, design patterns minimize 
misunderstandings to a minimum and establish consistency. Specifically, the use of known messages, signs, and symbols in a familiar structure enhances the user experience by promoting familiarity and intuitiveness as users encounter consistent interactions across applications and solutions. This not only reduces the learning curve for users, but also contributes to overall usability.
User input on a relatively small touch screen is a time-consuming task. As a result, the user's ability to remember the data entered decreases over time. The application should inform the user, if needed, about the saved information.

A mobile application can preset task hints for the necessary checkpoints. In fact, displaying task hints can provide guidance to users, especially new or inexperienced ones, by offering step-by-step instructions on how to perform certain actions or tasks within the app. On the other hand, task hints can highlight features or functionalities that users might not be aware of, thus improving the discoverability of those features.

Last but not least, while metadata aims to categorize and label data, it provides the essential semantics of the locally collected data by extending the limited description. In other words, the better the user understands the data, the greater the perceived usefulness in a given application.

\subsection{Errors}
A usability attribute termed as Errors reflects how well the user can perform the desired tasks without making errors, while the estimated user rate can be used to infer the simplicity of a system \cite{harrison2013usability}. Such information allows one to identify areas that are problematic for users and ultimately improve those areas in subsequent development iterations of the mobile application. 

On the other hand, users should make few errors while using an application, and if they do make errors, they should be able to easily recover from them \cite{ali2022mobile} by being provided with context-sensitive help and meaningful feedback when errors occur \cite{seffah2006usability}. By understanding the nature of these errors, it becomes possible to prevent them from occurring in next versions of the application \cite{parsazadeh2018framework}. 

That being said, Errors, as the mobile usability attribute, is manifested by the four application features, listed in Table \ref{tab:oper-errors}. 

\begin{table}[h]
\caption{Operationalization of the attribute Errors for mobile applications}
\label{tab:oper-errors}
\footnotesize
\begin{tabular}{|l|p{9cm}|p{2.5cm}|}
\hline
\textbf{Code}  & \textbf{Feature}   & \textbf{Related Work} \\ \hline
ER1 & ability to withdraw  the last performed action without losing the already input data &
\cite{franke2012mobile}, \cite{alvarado2018layered}, \cite{nunez2020model}       \\ \hline
ER2 & messages to prevent possible errors  &  \cite{inostroza2012usability}, \cite{kuparinen2013introducing} \\ \hline
ER3 & performing a wrong gesture does not result in application errors  & \cite{heuwing2015usability}, \cite{pushp2018privacyshield}, \cite{wu2020user}       \\ \hline
ER4 & verification of the correctness of the input data & \cite{schneider2008investigation}, \cite{qin2011usability}, \cite{george2018usability}  \\ \hline
\end{tabular}
\end{table}

Bearing in mind the different circumstances and associated limitations that a user may face, a mobile application should have the ability to skip an operation while preserving the user's input, such as form data or written notes. More generally, it refers to data persistence, which is the ability of an application to store and retrieve data after it is closed or the device is restarted. 

A message is a basic means of communication between an application and a user. A message can also assist a user by providing additional explanations or a possible solution. Considering current UX practices, error messages are implemented in a reactive manner, meaning that an application reacts to a specific situation (e.g. bad user password). In this sense, both error-preventive design and implemented data quality controls aim at reducing or even eliminating error-prone user behavior.

%Note that all four of these properties can be verified during application testing. The underlying goal, then, is to make the application more tolerant of user actions and to facilitate communication between applications.

\subsection{Cognitive Load}  %jest w ankiecie
Cognitive load refers to the amount of cognitive processing required by the user to use the application \cite{harrison2013usability}. 
In traditional usability studies a common assumption is that the user is performing only a single task and can therefore concentrate completely on that task. However, in a mobile context users will often be performing a second action in addition to using the mobile application \cite{deegan2015complex}. Thus, in this context, cognitive load includes mental load and mental effort, reflecting users intrinsic load which is a combination of extraneous load, and germane load \cite{sweller1998cognitive}. While the former refers to those mental resources which does not aid using mobile application, then the latter refers to the effort needed to use memory and intelligence to process information into schema.

Table \ref{tab:oper-cognitive-load} presents the manifesting variables that are used to operationalize the attribute of cognitive load. 

\begin{table}[h]
\caption{Mobile Cognitive Load Operationalization}
\label{tab:oper-cognitive-load}
\footnotesize
\begin{tabular}{|l|p{9cm}|p{2.5cm}|}
\hline
\textbf{Code}  & \textbf{Feature}  & \textbf{Related Work} \\ \hline
CL1 & application allows other activities to be performed at the same time & \cite{deegan2015complex}, \cite{weichbroth2020usability}, \cite{karczewska2021usability}   \\ \hline
CL2 & no user interaction required while the app is running in the background   & \cite{smith2017adaptive}, \cite{deegan2011usability}      \\ \hline
CL3 & ability to use other applications or device functions while the app is running in the background &   
\cite{harrison2013usability}, \cite{daud2023design}, \cite{deegan2014mobile}     \\ \hline
CL4 & ability to perform another activity without having to stop one already started  & \cite{harrison2013usability}, \cite{alasmari2020effect}, \cite{weichbroth2024usability}  \\ \hline
\end{tabular}
\end{table}

%CL4 & duration of mental effort during interaction with the application &  \cite{ibili2019assessing}, \cite{ilany2019mobile}, \cite{schobel2020measuring}    \\ \hline

As can be seen, cognitive load is operationalized by four indicators related to the level of user involvement required to perform tasks.
Here, intrinsic cognitive load is determined by the intrinsic nature of the information to be read and understood, more specifically, by the number of interacting information components that the performing task comprises. Note that, novice app users, with little or none any prior knowledge of the task, have to engage more mental effort, however, over time, as learning progresses, information components become incorporated (or chunked) into cognitive schemata. Therefore, the intrinsic cognitive load that is imposed by a task is much higher for novices than for more advanced app users. 

Note that extraneous cognitive load arises from the performance of tasks that require the user to engage in cognitive processes that do not directly contribute to the task by forcing the user to mentally integrate temporally or spatially separate but interrelated. For instance, a user may simultaneously check the physical route with the directions provided by a mobile navigation map due to current traffic jams or suddenly occurring adverse events. Therefore, extraneous cognitive is the result of confounding variables whose presence affects the perceived usability. 

In one perspective, researchers argue that germane load is synonymous with intrinsic load, since without intrinsic load there would be no germane load to facilitate the construction of new knowledge \cite{greenberg2023revisiting}, while in a more robust view, germane load is dependent on intrinsic load, it cannot become an autonomous source of cognitive load \cite{leppink2015evolution}. On the other hand, one can question the application capability to prioritize germane load, which in this case refers to the essential information that users need to acquire new knowledge effectively, unless such purpose is intentionally implemented and facilitated by specific app features. In this vein, for instance, one could point to the military services who typically take a great responsibility of the undertaken actions based on the information delivered by the mobile solutions. 

To summarize, considering the above arguments on the one hand, and since our goal is to operationalize the cognitive goal in a generic way and thus applicable to the broad types of mobile applications, the assigned manifest variables exclusively represent the intrinsic load. However, for contextual studies that aim to evaluate user behavior or perceptions under the intervention of the specific stimulus, other variables should be formulated and investigated.

\subsection{Understandability} %feedback and guidance 10 
Understandability refers to the capability of the mobile application to allow users to understand its application and to easily performs tasks \cite{ammar2019usability}. From the perspective of the interaction, understandability can be considered as the degree to which the application's messages are understood by the user. Typically, an individual message is the app reaction to user's performed action. In other words, such feedback can bring the form of graphics, text or a combination of both, and haptics, sounds and spoken messages as well. One can distinguish  three types of messages: input requests, status notifications, and error messages \cite{pfister2011affective}. 
Under specific circumstances, an app can guide a user by ad hoc delivering a brief documentation, context-aware help or a form-based wizards. 

Having said that, Table \ref{tab:oper-understand} presents the features that manifest mobile application understandability.

\begin{table}[h]
\caption{Operationalization of the attribute Understandability for mobile applications}
\label{tab:oper-understand}
\footnotesize
\begin{tabular}{|l|p{9cm}|p{2.5cm}|}
\hline
\textbf{Code}  & \textbf{Feature}  & \textbf{Related Work} \\ \hline
UY1 & availability of help or a user manual    & \cite{ahmad2021spiritual}, \cite{afrin2022usability}, \cite{castilla2023digital}  \\ \hline
UY2 & autocomplete feature with default values  &  \cite{chang2013improving}, \cite{piplani2018ict}, \cite{holstrom2020effects}      \\ \hline
UY3 & visual confirmation of the performed action  &  \cite{goumopoulos2017development}, \cite{perez2020evaluation}, \cite{tovide2022signsupport}     \\ \hline
UY4 & word completion feature  & \cite{agosti2003managing}, \cite{chittaro2007mobile}, \cite{bilal2018analyzing}   \\ \hline
UY5 & input data description  &  \cite{liang2017mobile}, \cite{ali2022mobile}    \\ \hline
\end{tabular}
\end{table}

Detailed help or tutorials allow users to solve common problems on their own, lowering the learning curve by making the application more accessible, especially for new users. Autocomplete speeds up data entry by predicting what the user is likely to type next, reducing the number of keystrokes required and making interaction more efficient, while default values provide quick options, minimizing user effort and making the application more user-friendly and helpful. In addition, new users may find it easier to learn and use the it when they see suggestions and default values, which can reduce the initial learning curve.

Visual confirmation provides immediate feedback, letting users know that their actions have been recognized and processed. This is essential for maintaining user confidence and ensuring they feel in control of the application, as well as guiding users through their interactions with the application and helping them understand what is happening at each step.

By design, word completion feature offers convenience and consistency. The former allows users to quickly select suggested words from a list, which is especially beneficial on mobile devices with smaller keyboards and screen sizes, while the latter ensures that commonly used words and phrases are entered consistently, which is important for maintaining data quality and governance.

Providing input data descriptions help users understand what information is required, in particular on the format, units, and type of data expected, which can be particularly helpful for complex or unfamiliar inputs. Note that in some cases, providing data descriptions is necessary to comply with regulatory standards and to ensure that all required information is collected accurately and completely. In general, descriptions help communicate the purpose and meaning of each data field, reducing ambiguity and improving overall communication between the application and the user.

Above properties being an application outputs, passively and actively support user confidence during tasks’ performance. The actions undertaken should neither distract the user nor obscure the current display. Providing appropriate feedback and guidance leads to a better understanding of the application's functionality.

\subsection{Operability} %W ankiecie jest visibility
Operability is the ability of the mobile application to allow the user to operate and control it in different contexts of use \cite{weichbroth2024usability}. 

More specifically, when visual on-screen objects are clearly visible and easily accessible within the interface, users can more effectively navigate the application and perform tasks. In this line of thinking, visibility seems to be a closely related concept understood in terms of perceived affordances or signifiers \cite{norman2010gestural}. An affordance is the design aspect of an object that suggests how the object should be used \cite{mcgrenere2000affordances}, while signifiers are visual cues that indicate the affordances of an application.

Due to the physical limitations of mobile devices, the complexity of graphics and icons has been reduced to a minimum, while the elements of the user interface have to be of a size that allows a user to easily manipulate them by performing certain touch gestures, on the one hand, and still have to be understood by preserving the true meaning. 

Table \ref{tab:oper-operability} shows the variables used to operationalize the operability of the mobile application.

\begin{table}[h]
\caption{Operationalization of the attribute Operability for mobile applications}
\label{tab:oper-operability}
\footnotesize
\begin{tabular}{|l|p{9cm}|p{2.7cm}|}
\hline
\textbf{Code}  & \textbf{Feature}  & \textbf{Related Work} \\ \hline
OP1 & location of text boxes and buttons & \cite{kangas2005applying}, \cite{nguyen2015reverse}, \cite{natarajan2018p2a} \\ \hline
OP2 & size of text boxes and buttons  & \cite{kangas2005applying}, \cite{deniz2019comparison}       \\ \hline
OP3 & language used to describe text boxes  &  \cite{kangas2005applying}, \cite{kim2022speak}, \cite{wen2023droidbot}   \\ \hline
OP4 & current progress of the task being performed by the application  & \cite{mansar2012usability},  \cite{moroyoqui2022smartasko}, \cite{nakagawa2022implementation}  \\ \hline
\end{tabular}
\end{table}

The location of text boxes and buttons in mobile applications aims to establish a clear visual hierarchy, maintain contextual relevance, and guide user focus. Appropriate size allows to organize input fields and buttons in an accessible and intuitive sequence, matching the user's natural flow of interaction. 
Using language relevant to the specific context of each text box provides a better understanding of the purpose of each input field by explaining the context necessary for accurate data entry. 
In the case of complex task performance, displaying the current progress of tasks informs the user that the application is working on their request and gives them a sense of the remaining work needed to complete the task. On the other hand, a real-time update progress message assures the user that their input has been acknowledged and is being processed by the application.

In practice, the elements on the screen must be adequately aligned and contrasted \cite{cunha2013heuristic}. Taking into account physical limitations of the mobile devices applying these properties leads to an increase in the effectiveness of the user in completing the task.

\section{Methodology}
\label{sec:methodology}
After establishing the theoretical foundation through the conceptualization and operationalization of mobile usability factors, the following research question is posed: What factors, and to what extent, directly influence the perceived usability of mobile applications?
To answer this question, we conducted a survey, following and adapting the guidelines developed by Bennett et al. \cite{bennett2011reporting} and Passmore et al. \cite{passmore2002guidelines}. We designed a questionnaire with scalable responses to collect the necessary data and elicit unambiguous responses that could be quantified and compared across a sample of respondents.

\subsection{Survey design}
The survey was divided into three parts. 
The first part began with an introductory statement and a set of standard demographic questions to identify respondents age, gender, education and professional experience in number of years. 
In second part, respondents were asked to select the types of mobile devices (tablet, e-book reader, and smartphone) and their respective operating systems (Android OS, Apple iOS, Windows OS, BlackBerry OSx, and other) and the corresponding built-in operating system, utilizing for at least 3 months.
The third part of the survey contained a list of 46 features to rate the extent to which they affect the perceived usability of the mobile applications usability, based on his/her experience. In this regard, we use the measurement scale which was divided into five categories: (1) very weak, (2) weak, (3) moderate, (4) strong and (5) very strong.

\subsection{Data gathering and pre-processing}
Data gathering was carried out both on the paper and electrically.
In the case of the former type was submitted in case of the lack of the desktop computers since the questionnaire web page was not optimized for mobile devices, while in case of the latte type, a link was sent by the email to the people who agreed to participate in the study. We used a well-known and free of charge survey administration software tool to collect the data.

The study period was 6 months, i.e. March-August 2018, using convenience sampling method \cite{emerson2015convenience}. 
It means that data was collected from the population members who were conveniently available to participate in the study. This method involves getting respondents wherever the circumstances and willingness to participate occurred. Therefore, no inclusion (exclusion) criteria were identified and applied prior to the selection process. It is worth noting that respondents did not receive any form of reward for completing the survey.

All paper-based responds were manually input into the database. The gathered data was exported in the plain text format, and processed in the spreadsheet desktop application.
Next the merged dataset was analyzed to exclude unreliable respondents who have selected: (a) all possible answers in the section 2, OR in the section 3, OR (b) the same answer in the section 6. In total, six respondents were excluded, since they met at least one of the three above conditions. The table \ref{tab:survey-summary} provides a summary of the survey in this scope.

\begin{table}[h]
\centering
\small
\begin{tabular}{|l|l|l|}
\hline
\textbf{Observations} & \textbf{Volume}   & \textbf{Share}     \\ \hline
Valid        & 838 & 99.30\% \\ \hline
Excluded     & 6   & 0.71\%  \\ \hline
All          & 844 & 100\% \\ \hline
\end{tabular}
\caption{\label{tab:survey-summary}The summary of collected data, after initial pruning.}
\end{table}

As one can observe, the valid number of respondents was 838, therefore the exclusion rate was about 0.71\%, which can be considered very low.

\subsection{Sample Characteristics}
The mean age of 838 participants who gave their age was 25.5 years (SD 8.00). The youngest respondent was 16 while the oldest was 65. The sample was divided into six age groups, depicted by Table \ref{tab:sample-age}.

\begin{table}[h]
\centering
\small
\begin{tabular}{|l|l|l|}
\hline
\textbf{Age group} & \textbf{Volume}  & \textbf{Share}     \\ \hline
19 or less      & 106   & 12,65\% \\ \hline
20--24          & 425   & 50,72\% \\ \hline
25--29          & 111   & 13,25\% \\ \hline
30--34          & 91    & 10,86\% \\ \hline
35--39          & 44    & 5,25\% \\ \hline
40 or more      & 61    & 7,28\%  \\ \hline
\textbf{All}    & \textbf{838}   & \textbf{100\%} \\ \hline
\end{tabular}
\caption{\label{tab:sample-age}The age distribution of the respondents.}
\end{table}

As once can notice, over half of the study participants (50.72\%) ranged from 20 to 24 years old. Not surprisingly, the larger part of the remaining (25.9\%) concerns the older (13.25\%) and the younger group (12.65\%). The rest (23.38\%) includes the groups of subject at the age of 30 and greater.

%gender
The sample consists of 354 (42.2\%) women (\female)  and 484 (57.8\%) men (\male).
%education level
The basic education was declared by 68 (8.1\%) respondents (28 women and 40 men), secondary by 483 (57.6\%, 189 women and 294 men), while higher education by 287 (34.2\%, 137 women and 150 men). 
%proffesional experience 

No work experience was reported by 335 respondents (40\%, 134 \female, and 201 \male), 
one year by 75 (8.9\%, 28 \female, and 49 \male), two years by 91 (10.9\%, 35 \female, and 56 \male), and three years or more by 337 (40.2\%, 157 \female, and 180 \male). 

For another question related to the mobile devices and embedded operating systems used, we divided the survey sample into two mutually exclusive subsets based on the gender of the respondents. In this regard, the details in case of women and men are provided by Table \ref{tab:device-OS-w} and Table \ref{tab:device-OS-m}, respectively.

\begin{table}[h]
\centering
\small
\begin{tabular}{|l|l|l|l|l|l|}
\hline
\textbf{Device} \textbackslash \textbf{OS} & \textbf{Android} & \textbf{iOS} & \textbf{Windows} & \textbf{BlackBerry} & \textbf{Other} \\ \hline
Tablet        & 98 (27.68\%)   & 26 (7.34\%)   & 25 (7.06\%) & 0 (0\%) & 3 (0.85\%)     \\ \hline
Ebook Reader  & 24  (6.78\%)   & 9   (2.54\%)   & 3 (0.85\%)  & 1 (0.28\%)  & 24 (6.78\%)    \\ \hline
Smartphone    & 249  (70.34\%)  & 90  (25.42\%) & 18 (5.08\%) & 0  (0\%)    & 0  (0\%)   \\ \hline
\end{tabular}
\caption{\label{tab:device-OS-w}The device and built-in operating system used by women.}
\end{table}

Among women (Table \ref{tab:device-OS-w}), the most popular mobile device was a smartphone, with Android (70.34\%), iOS (25.42\%) and Windows (5.08\%) on board. The second place took Tablet, with Android (27.68\%), iOS, (7.34\%) and Windows (7.06\%), as the host operating systems. The ebook reader proved to be the least popular, where only Android share was significant (6.78\%), as the share of the two remaining OS did not exceed five percent, except those with no specified operating system (6.78\%).

\begin{table}[h]
\centering
\small
\begin{tabular}{|l|l|l|l|l|l|}
\hline
\textbf{Device} \textbackslash  \textbf{OS} & \textbf{Android} & \textbf{iOS} & \textbf{Windows} & \textbf{BlackBerry} & \textbf{Other} \\ \hline
Tablet        & 152 (31.40\%)  & 47 (9.71\%) & 27 (5.58\%) & 0 (0\%) & 9 (1.86\%)     \\ \hline
Ebook Reader  & 25  (5.17\%) & 6  (1.24\%) & 7 (1.45\%) & 0 (0\%) & 48 (9.92\%)   \\ \hline
Smartphone    & 375 (77.48\%) & 106 (21.9\%) & 20 (4.13\%) & 0 (0\%) & 0 (0\%) \\ \hline
\end{tabular}
\caption{\label{tab:device-OS-m}The device and built-in operating system used by men}
\end{table}
When considering men (Table \ref{tab:device-OS-m}), and comparing with women, the results are similar. Again, the most popular mobile device was smartphone, running under Android (77.48\%), iOS (31.04\%) and Windows (5.17\%), followed by tablet with Android (31.40\%), iOS (9.71\%), and Windows (5.58\%), along with ebook reader on the non-specified OS (9.92\%) and Android (5.17\%). 

Interestingly, 20 respondents (7 \female, and 13 \male) declared not using smartphones at all, whereas 34 (7 \female, and 27 \male) reported using two smartphones, and 4 (2 \female, and 2 \male) even three smartphones at the same time.

\section{Results}
\label{sec:Results}
%factors evaluation results
Quantitative data from the questionnaire were analyzed using a well-known commercial spreadsheet software package along with jamovi, a free and open source software for data analysis and statistical tests. 
Note that Jamovi provides easy-to-use statistical software with a point-and-click interface and supports a wide range of built-in statistical tests, real-time output, data visualization, and R integration. In addition, it is lightweight, cross-platform compatible, and regularly updated, making it a powerful statistical tool that has been used in numerous studies to date.

\begin{figure}[ht] 
    \centering 
    %\includesvg[svgpath=./, width=0.95\textwidth]{analysis-procedure.svg} 
    \includegraphics[scale=0.65]{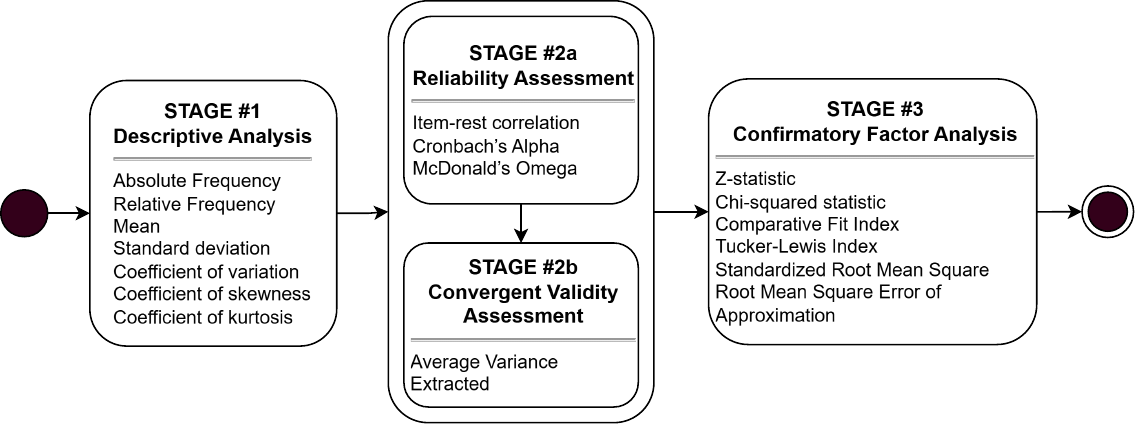}
    \caption{Analysis procedure.} 
    \label{fig:analysis-procedure} 
\end{figure}

The analysis of the collected data and the subsequent presentation of the results have been divided into three stages, as shown in Figure~\ref{fig:analysis-procedure}. Additionally, we conclude the analysis by highlighting the key findings at the end.

In the first stage (descriptive analysis), analyses were performed using descriptive statistics such as: absolute and relative frequencies, mean (\textit{M}), standard deviation (\textit{SD}), coefficient of variation (\textit{CV}), coefficient of skewness (\textit{Sk}), and coefficient of kurtosis ($\kappa$). 
In the second stage (reliability and convergent validity assessment), we considered item-rest correlation to assess the association of a particular item  with the rest of the scale. We also calculated Cronbach's $\alpha$ and McDonald's $\omega$ to measure and evaluate internal consistency. Afterwards, we used Average Variance Extracted (AVE) in terms of measuring and evaluating convergent validity. 
In the third stage, we performed confirmatory factor analysis to test and evaluate a predefined model of how observed variables relate to underlying latent factors. In addition, we used indices such as the chi-squared statistic and goodness-of-fit indices (CFI and TLI), as well as error statistics (SRMR and RMSEA), to determine if the hypothesized theoretical model is a reasonable representation that can be used to draw inferences.
Finally, we conclude the findings with a brief summary.

A frequency analysis is the first step in results evaluation. In this extent, Table \ref{tab:all-features-shares} provides the results of calculated absolute and relative frequency of scores, showing the level of the features impact on perceived usability, given by the study respondents. As one can notice, out of the 32 features evaluated individually and while considering the highest calculated relative share, thirteen received a high score (4), twelve a moderate score (3), and seven a very high score (5). In addition, further analysis of the distribution of score values allows us to conclude that in case of all features the relative share of very low scores (1) was the lowest, followed by the low scores (2). 

\begin{table}[]
\caption{Calculated absolute and relative frequency of scores reflecting how much certain features affect perceived usability.}
\label{tab:all-features-shares}
\footnotesize
\begin{tabular}{|l|l|l|l|l|l|}
\hline
\textbf{ID} / \textbf{Score} & \textbf{very weak (1)} & \textbf{weak (2)} & \textbf{moderate (3)}  & \textbf{high (4)} & \textbf{very high (5)} \\ \hline
ES1               & 33 (3.93\%)   & 90 (10.73\%)  & 280 (33.41\%) & 278 (33.17\%) & 157 (18.73\%) \\ \hline
ES2               & 39 (4.65\%)   & 94 (11.21\%)  & 305 (36.39\%) & 271 (32.33\%) & 129 (15.39\%) \\ \hline
ES3               & 37 (4.41\%)   & 62 (7.39\%)   & 251 (29.95\%) & 274 (32.69\%) & 214 (25.53\%) \\ \hline
ES4               & 32 (3.81\%)   & 54 (6.44\%)   & 247 (29.47\%) & 309 (36.87\%) & 196 (23.38\%) \\ \hline
EY1               & 26 (3.1\%)    & 65 (7.75\%)   & 160 (19.09\%) & 235 (28.04\%) & 352 (42\%)    \\ \hline
EY2               & 22 (2.62\%)   & 54 (6.44\%)   & 157 (18.73\%) & 237 (28.28\%) & 368 (43.91\%) \\ \hline
EY3               & 22 (2.62\%)   & 45 (5.36\%)   & 146 (17.42\%) & 176 (21\%)    & 449 (53.57\%) \\ \hline
EY4               & 20 (2.38\%)   & 51 (6.08\%)   & 146 (17.42\%) & 251 (29.95\%) & 370 (44.15\%) \\ \hline
LY1               & 18 (2.14\%)   & 34 (4.05\%)   & 225 (26.84\%) & 283 (33.77\%) & 278 (33.17\%) \\ \hline
LY2               & 39 (4.65\%)   & 83 (9.9\%)    & 248 (29.59\%) & 256 (30.54\%) & 212 (25.29\%) \\ \hline
LY3               & 57 (6.8\%)    & 127 (15.15\%) & 277 (33.05\%) & 246 (29.35\%) & 131 (15.63\%) \\ \hline
MY1               & 38 (4.53\%)   & 62 (7.39\%)   & 275 (32.81\%) & 300 (35.79\%) & 163 (19.45\%) \\ \hline
MY2               & 35 (4.17\%)   & 74 (8.83\%)   & 271 (32.33\%) & 284 (33.89\%) & 174 (20.76\%) \\ \hline
MY3               & 37 (4.41\%)   & 89 (10.62\%)  & 340 (40.57\%) & 266 (31.74\%) & 106 (12.64\%) \\ \hline
MY4               & 46 (5.48\%)   & 97 (11.57\%)  & 351 (41.88\%) & 244 (29.11\%) & 100 (11.93\%) \\ \hline
ER1               & 28 (3.34\%)   & 59 (7.04\%)   & 214 (25.53\%) & 236 (28.16\%) & 301 (35.91\%) \\ \hline
ER2               & 32 (3.81\%)   & 79 (9.42\%)   & 251 (29.95\%) & 304 (36.27\%) & 172 (20.52\%) \\ \hline
ER3               & 24 (2.86\%)   & 68 (8.11\%)   & 245 (29.23\%) & 283 (33.77\%) & 218 (26.01\%) \\ \hline
ER4               & 33 (3.93\%)   & 65 (7.75\%)   & 238 (28.4\%)  & 307 (36.63\%) & 195 (23.26\%) \\ \hline
CL1               & 36 (4.29\%)   & 60 (7.15\%)   & 228 (27.2\%)  & 248 (29.59\%) & 266 (31.74\%) \\ \hline
CL2               & 34 (4.05\%)   & 85 (10.14\%)  & 293 (34.96\%) & 266 (31.74\%) & 160 (19.09\%) \\ \hline
CL3               & 35 (4.17\%)   & 51 (6.08\%)   & 241 (28.75\%) & 258 (30.78\%) & 253 (30.19\%) \\ \hline
CL4               & 35 (4.17\%)   & 61 (7.27\%)   & 189 (22.55\%) & 255 (30.42\%) & 298 (35.56\%) \\ \hline
UY1               & 63 (7.51\%)   & 131 (15.63\%) & 293 (34.96\%) & 235 (28.04\%) & 116 (13.84\%) \\ \hline
UY2               & 43 (5.13\%)   & 112 (13.36\%) & 321 (38.3\%)  & 254 (30.31\%) & 108 (12.88\%) \\ \hline
UY3               & 36 (4.29\%)   & 87 (10.38\%)  & 292 (34.84\%) & 297 (35.44\%) & 126 (15.03\%) \\ \hline
UY4               & 59 (7.04\%)   & 116 (13.84\%) & 320 (38.18\%) & 252 (30.07\%) & 91 (10.85\%)  \\ \hline
UY5               & 59 (7.04\%)   & 114 (13.6\%)  & 343 (40.93\%) & 222 (26.49\%) & 100 (11.93\%) \\ \hline
OP1               & 19 (2.26\%)   & 47 (5.6\%)    & 224 (26.73\%) & 334 (39.85\%) & 214 (25.53\%) \\ \hline
OP2               & 15 (1.78\%)   & 65 (7.75\%)   & 307 (36.63\%) & 298 (35.56\%) & 153 (18.25\%) \\ \hline
OP3               & 28 (3.34\%)   & 88 (10.5\%)   & 300 (35.79\%) & 276 (32.93\%) & 146 (17.42\%) \\ \hline
OP4               & 23 (2.74\%)   & 93 (11.09\%)  & 288 (34.36\%) & 313 (37.35\%) & 121 (14.43\%) \\ \hline
\end{tabular}
\end{table}

In this line of thinking, the importance of the usability factors is also evident from the results of the descriptive statistics calculated. Reading Table \ref{tab:ds-all-features} across the rows, and taking into account the 5-point scale used, it seems clear that on average three features were considered highly important and the rest moderately important. A standard deviation, a measure of how dispersed the data is in relation to the mean, is in the range between 0.9312 and 1.1147. In addition, its relative size compared to the mean is in the range of 24.88\% and 34.12\%. While a rule of thumb says that a coefficient of variation (\textit{CV}) of 30\% or less is a desirable level, slightly higher \textit{CV}s are acceptable in survey research due to the natural, inherent variability in human perception.

\begin{table}[]
\small
\caption{Summary of descriptive statistics calculated for mobile usability features.}
\label{tab:ds-all-features}
\begin{tabular}{|l|l|l|l|l|l|}
\hline
\textbf{ID} / \textbf{Statistics}  & \textbf{M} & \textbf{SD}  & \textbf{CV} & \textbf{Sk} & \textbf{$\kappa$}     \\ \hline
ES1 & 3.5203  & 1.0376 & 0.2948 & -0.3694 & -0.3116  \\ \hline
ES2 & 3.4260  & 1.0282 & 0.3001 & -0.3291 & -0.2431  \\ \hline
ES3 & 3.6754  & 1.0697 & 0.2910 & -0.5585 & -0.1631  \\ \hline
ES4 & 3.6957  & 1.0192 & 0.2758 & -0.5993 & 0.0720   \\ \hline
EY1 & 3.9809  & 1.0962 & 0.2754 & -0.8842 & -0.0256  \\ \hline
EY2 & 4.0442  & 1.0583 & 0.2617 & -0.9491 & 0.1802   \\ \hline
EY3 & 4.1754  & 1.0638 & 0.2548 & -1.1480 & 0.4949   \\ \hline
EY4 & 4.0740  & 1.0343 & 0.2539 & -0.9991 & 0.3494   \\ \hline
LY1 & 3.9177  & 0.9748 & 0.2488 & -0.6569 & 0.0616   \\ \hline
LY2 & 3.6193  & 1.1048 & 0.3052 & -0.4797 & -0.4107  \\ \hline
LY3 & 3.3186  & 1.1147 & 0.3359 & -0.2682 & -0.5676  \\ \hline
ES4 & 3.6957  & 1.0192 & 0.2758 & -0.5993 & 0.0720   \\ \hline
MY1 & 3.5823  & 1.0264 & 0.2865 & -0.5183 & 0.0005   \\ \hline
MY2 & 3.5823  & 1.0426 & 0.2910 & -0.4575 & -0.1970  \\ \hline
MY3 & 3.3759  & 0.9829 & 0.2911 & -0.2867 & -0.0677  \\ \hline
MY4 & 3.3043  & 1.0062 & 0.3045 & -0.2565 & -0.1174  \\ \hline
ER1 & 3.8628  & 1.0860 & 0.2812 & -0.6847 & -0.2315  \\ \hline
ER2 & 3.6026  & 1.0339 & 0.2870 & -0.4940 & -0.1820  \\ \hline
ER3 & 3.7196  & 1.0283 & 0.2765 & -0.5021 & -0.2551  \\ \hline
ER4 & 3.6754  & 1.0379 & 0.2824 & -0.5862 & -0.0495  \\ \hline
CL1 & 3.7733  & 1.1012 & 0.2918 & -0.6403 & -0.2222  \\ \hline
CL2 & 3.5167  & 1.0388 & 0.2954 & -0.3491 & -0.2998  \\ \hline
CL3 & 3.7673  & 1.0752 & 0.2854 & -0.6310 & -0.1129  \\ \hline
CL4 & 3.8592  & 1.1090 & 0.2874 & -0.7787 & -0.0865  \\ \hline
UY1 & 3.2506  & 1.1090 & 0.3412 & -0.2275 & -0.5456  \\ \hline
UY2 & 3.3246  & 1.0264 & 0.3087 & -0.2495 & -0.2914  \\ \hline
UY3 & 3.4654  & 1.0080 & 0.2909 & -0.3998 & -0.1345  \\ \hline
UY4 & 3.2387  & 1.0486 & 0.3238 & -0.2891 & -0.3068  \\ \hline
UY5 & 3.2267  & 1.0535 & 0.3265 & -0.2106 & -0.2970  \\ \hline
OP1 & 3.8079  & 0.9567 & 0.2512 & -0.6176 & 0.1656   \\ \hline
OP2 & 3.6074  & 0.9312 & 0.2581 & -0.2572 & -0.2081  \\ \hline
OP3 & 3.5060  & 1.0050 & 0.2867 & -0.3066 & -0.2742  \\ \hline
OP4 & 3.4964  & 0.9626 & 0.2753 & -0.3403 & -0.1852  \\ \hline
\end{tabular}
\end{table}

We used Cronbach's alpha ($\alpha$) and McDonald's omega ($\omega$) to measure the internal consistency of each usability factor, along with a set of features assembled. In this regard, Table \ref{tab:reliability-all-items} presents the values of both calculated measures, along with item-rest correlation, used to evaluate the association of the feature with the total score on the other features. 
The general rule is that $\alpha$ or $\omega$ values between 0.6 and 0.90 indicate acceptable internal consistency in exploratory research. Coefficient values below 0.5 are usually unacceptable, while those between 0.5 and 0.6 can still be considered sufficient. In our study, although only one indicator (LY2) is 0.509, while the others range from 0.605 to 0.897, and all item-rest correlations are greater than 0.3, internal consistency seems to be achieved. 

\begin{table}[]
\caption{Summary of features Reliability statistics.}
\label{tab:reliability-all-items}
\small
\begin{tabular}{|l|l|l|l|}
\hline
\textbf{ID} / \textbf{Statistics}  & \textbf{Item-rest correlation} & \textbf{Cronbach's $\alpha$} & \textbf{McDonald's $\omega$} \\ \hline
ES1 & 0.470                 & 0.804        & 0.808        \\ \hline
ES2 & 0.634                 & 0.725        & 0.745        \\ \hline
ES3 & 0.662                 & 0.709        & 0.712        \\ \hline
ES4 & 0.650                 & 0.717        & 0.734        \\ \hline
EY1 & 0.793                 & 0.897        & 0.898        \\ \hline
EY2 & 0.840                 & 0.881        & 0.883        \\ \hline
EY3 & 0.810                 & 0.891        & 0.893        \\ \hline
EY4 & 0.792                 & 0.897        & 0.899        \\ \hline
LY1 & 0.479                 & 0.605        & 0.605        \\ \hline
LY2 & 0.548                 & 0.509        & 0.512        \\ \hline
LY3 & 0.457                 & 0.634        & 0.637        \\ \hline
MY1 & 0.710                 & 0.786        & 0.788        \\ \hline
MY2 & 0.722                 & 0.780        & 0.784        \\ \hline
MY3 & 0.667                 & 0.805        & 0.816        \\ \hline
MY4 & 0.611                 & 0.828        & 0.834        \\ \hline
ER1 & 0.684                 & 0.811        & 0.812        \\ \hline
ER2 & 0.710                 & 0.800        & 0.804        \\ \hline
ER3 & 0.710                 & 0.800        & 0.803        \\ \hline
ER4 & 0.652                 & 0.825        & 0.825        \\ \hline
CL1 & 0.766                 & 0.812        & 0.815        \\ \hline
CL2 & 0.650                 & 0.858        & 0.858        \\ \hline
CL3 & 0.735                 & 0.825        & 0.828        \\ \hline
CL4 & 0.727                 & 0.828        & 0.831        \\ \hline
UY1 & 0.585                 & 0.841        & 0.842        \\ \hline
UY2 & 0.713                 & 0.806        & 0.810        \\ \hline
UY3 & 0.697                 & 0.811        & 0.814        \\ \hline
UY4 & 0.671                 & 0.817        & 0.820        \\ \hline
UY5 & 0.646                 & 0.824        & 0.828        \\ \hline
OP1 & 0.638                 & 0.767        & 0.771        \\ \hline
OP2 & 0.696                 & 0.741        & 0.742        \\ \hline
OP3 & 0.641                 & 0.767        & 0.777        \\ \hline
OP4 & 0.573                 & 0.797        & 0.806        \\ \hline
\end{tabular}
\end{table}

\begin{figure}[htbp]
    \centering
    \small
    \subfloat[Effectiveness]{{\includegraphics[width=0.45\textwidth, height=4.8cm]{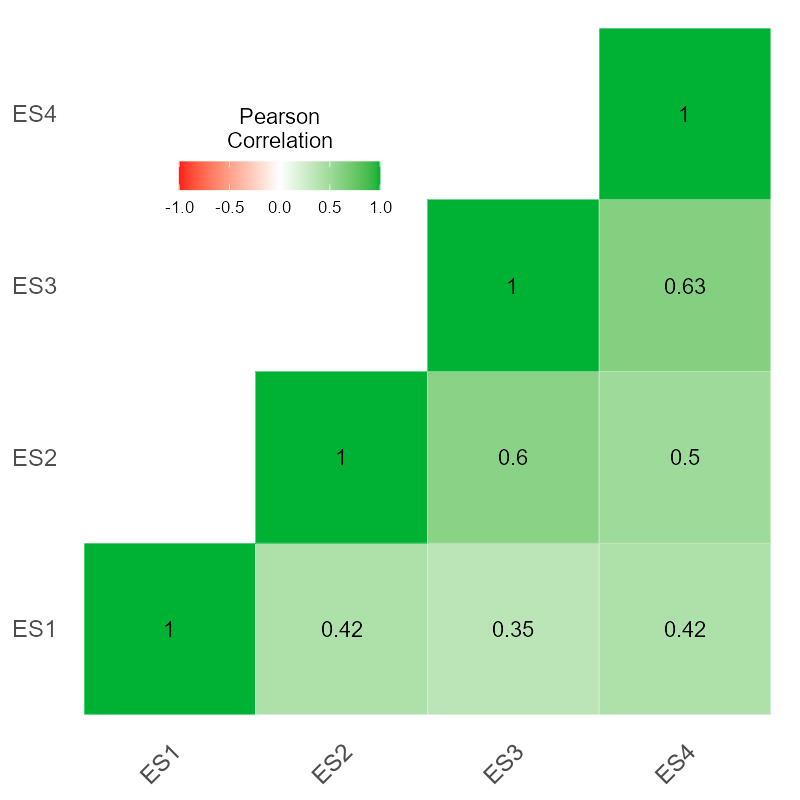}}}
    \subfloat[Efficiency]{{\includegraphics[width=0.45\textwidth, height=4.6cm]{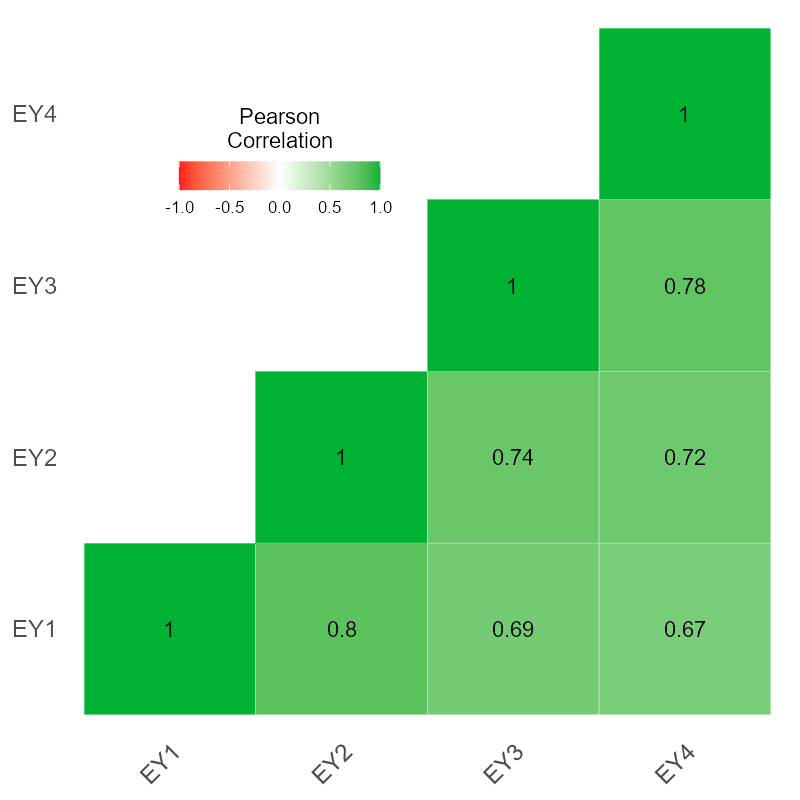}}}
    \hfill
    \subfloat[Learnability]{{\includegraphics[width=0.45\textwidth, height=4.6cm]{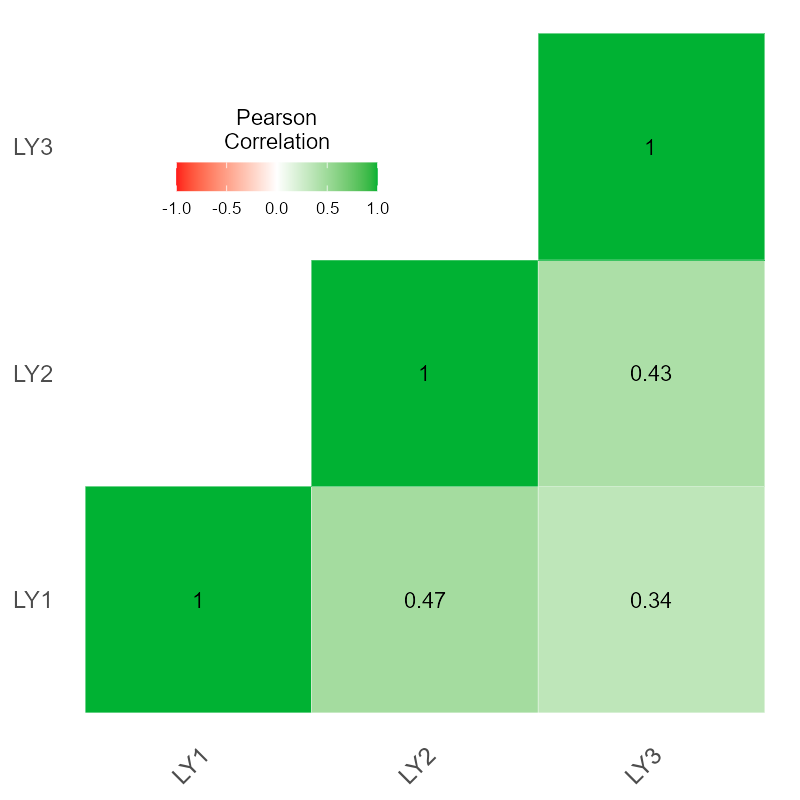}}}
    \subfloat[Memorability]{{\includegraphics[width=0.45\textwidth, height=4.6cm]{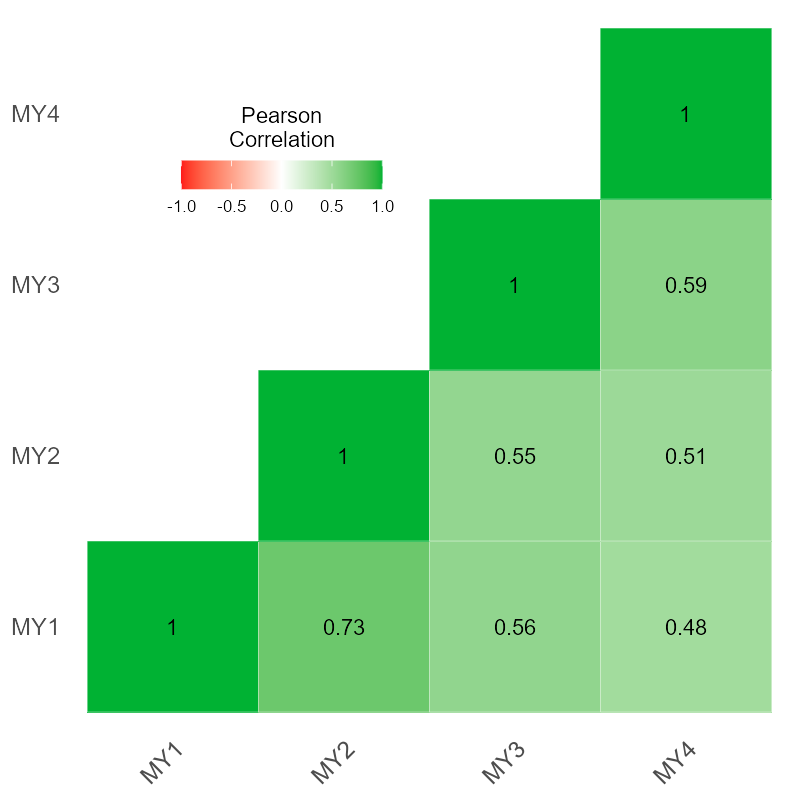}}}
    \hfill
    \subfloat[Errors]{{\includegraphics[width=0.45\textwidth, height=4.6cm]{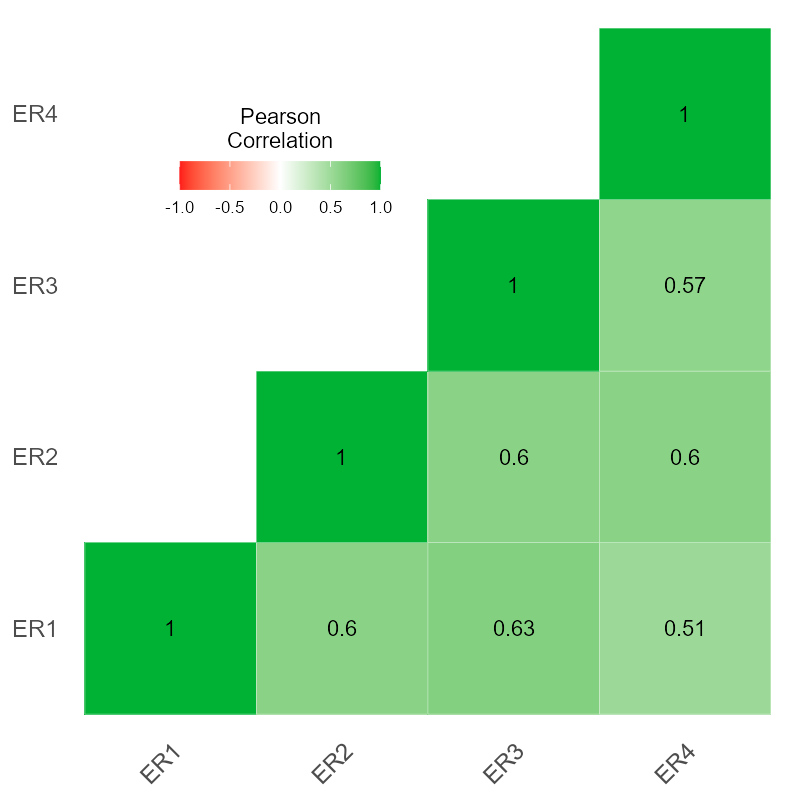}}}
    \subfloat[Cognitive Load]{{\includegraphics[width=0.45\textwidth, height=4.6cm]{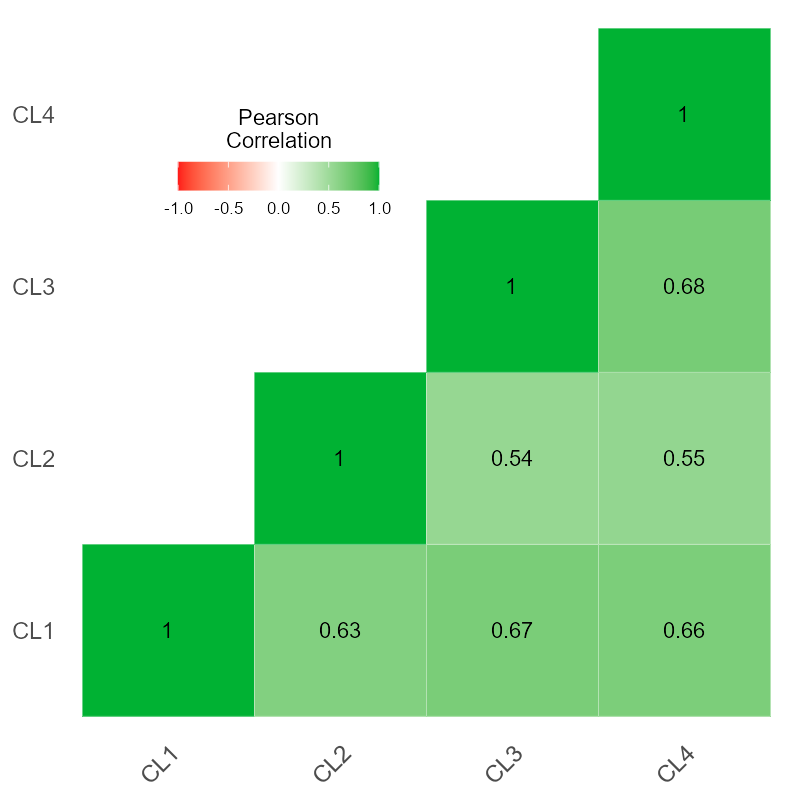}}}
    \hfill
    \subfloat[Understandability]{{\includegraphics[width=0.45\textwidth, height=4.6cm]{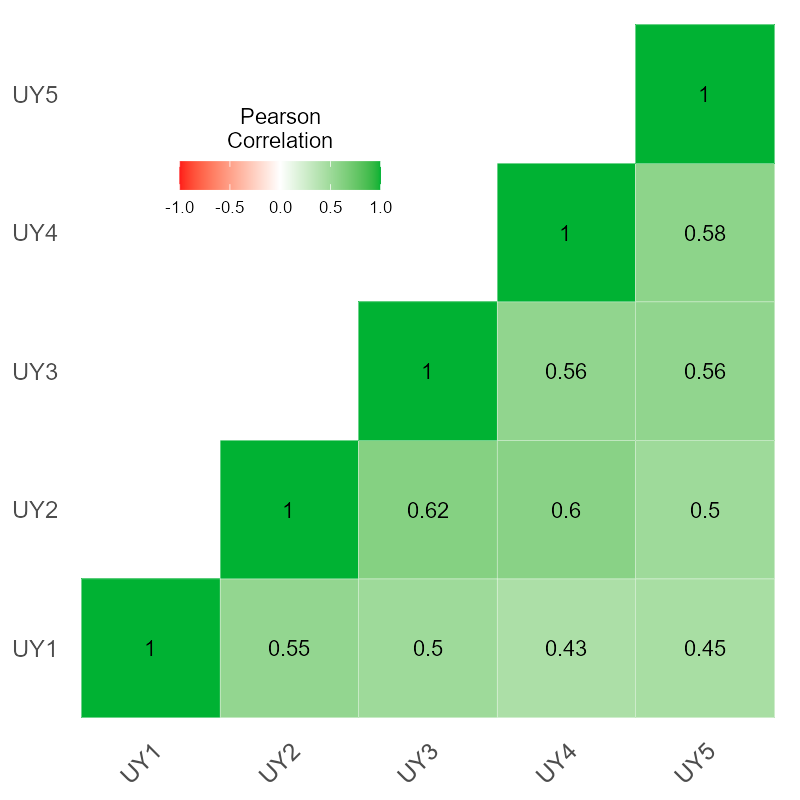}}}
    \subfloat[Operability]{{\includegraphics[width=0.45\textwidth, height=4.6cm]{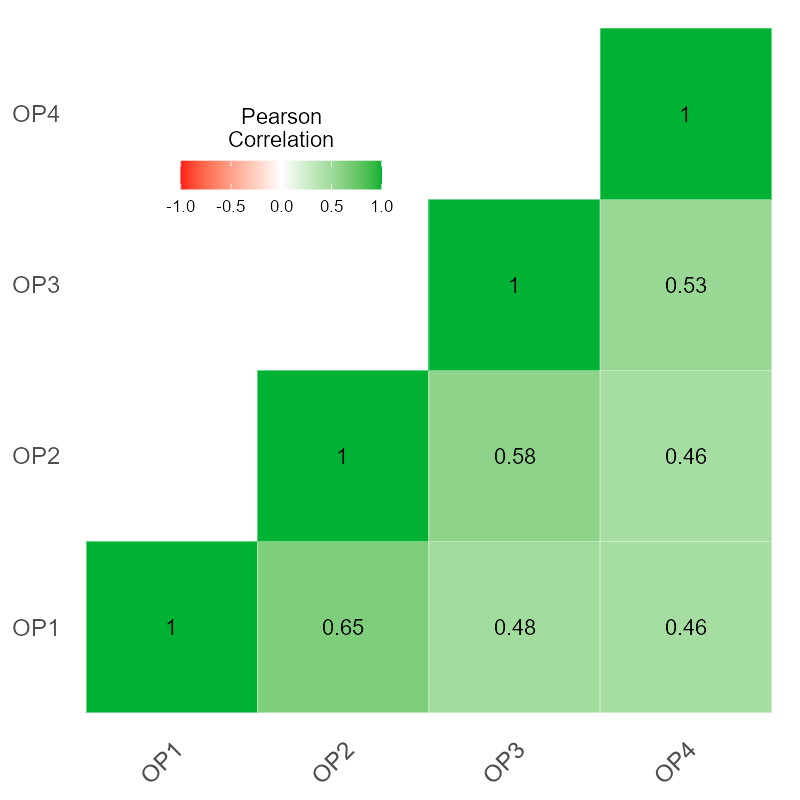}}} 
    \caption{Correlation heatmaps of mobile usability features.}
    \label{fig:heatmaps}
\end{figure}

Figure \ref{fig:heatmaps} shows the correlation between features for each individual factor, depicted in the form of a heat map. By analyzing the values, one can conclude that the strength of the correlations varies greatly between the factors, ranging from weak (less than 0.4) to very strong (0.8 or more). 

On the other hand, the lack of opposite, negative correlations suggests that there are no opposing effects of individual features on the perceived usability of mobile applications. Since all features exhibit a positive correlation, it can be argued that the test results indicate reliability.

While our research prespecifies the nature of the latent variables, we then need to test whether the predetermined factors adequately explain the intercorrelations among the items. In such a scenario, confirmatory factor analysis is a commonly used technique.

By definition, the validity of the measurement model is assessed by comparing the theoretical model to the actual data by examining the factor loadings. In general, a value of 0.7 or higher indicates that the factor adequately captures the variance of that variable. In other words, an adequate factor loading has a threshold of 0.7 or higher. Consequently, factor loadings help determine the importance and contribution of each variable to a factor. However, for a newly developed item, the factor loading for each item should be greater than 0.6.

Based on the results presented in Table~\ref{tab:CFA-factor-loadings}, none of the factor loadings for any factor is less than 0.6. Therefore, we can conclude that all the factor loadings have reached the acceptable level. However, ES1 (0.612), LY1 (0.696), MY4 (0.677), and OP4 (0.620) are less than 0.7. The standard error (SE) for all estimated factor loadings is less than 0.04, except for LY4 (0.0403), indicating high reliability of the mean.

\begin{table}[]
\caption{Summary of the Confirmatory Factor Analysis.}
\label{tab:CFA-factor-loadings}
\small
\begin{tabular}{|l|l|c|c|c|c|c|}
\hline
\textbf{Factor}  & \textbf{Indicator} & \textbf{Estimate} & \textbf{SE} & \textbf{Z} & \textbf{p}  & \textbf{Stand. Estimate} \\ \hline
\multirow{4}{*}{Efficiency}        & EY1       & 0.927    & 0.0312 & 29.7 & $<$.001 & 0.846           \\ \cline{2-7} 
                                   & EY2       & 0.938    & 0.0293 & 32.0 & $<$.001 & 0.887           \\ \cline{2-7} 
                                   & EY3       & 0.911    & 0.0301 & 30.2 & $<$.001 & 0.857           \\ \cline{2-7} 
                                   & EY4       & 0.865    & 0.0297 & 29.2 & $<$.001 & 0.837           \\ \hline
\multirow{4}{*}{Cognitive Load}    & CL1       & 0.917    & 0.0322 & 28.5 & $<$.001 & 0.833           \\ \cline{2-7} 
                                   & CL2       & 0.745    & 0.0325 & 22.9 & $<$.001 & 0.717           \\ \cline{2-7} 
                                   & CL3       & 0.874    & 0.0318 & 27.4 & $<$.001 & 0.813           \\ \cline{2-7} 
                                   & CL4       & 0.882    & 0.0332 & 26.6 & $<$.001 & 0.795           \\ \hline
\multirow{4}{*}{Errors}            & ER1       & 0.871    & 0.0326 & 26.7 & $<$.001 & 0.802           \\ \cline{2-7} 
                                   & ER2       & 0.778    & 0.0319 & 24.4 & $<$.001 & 0.753           \\ \cline{2-7} 
                                   & ER3       & 0.812    & 0.0310 & 26.2 & $<$.001 & 0.790           \\ \cline{2-7} 
                                   & ER4       & 0.739    & 0.0328 & 22.6 & $<$.001 & 0.713           \\ \hline
\multirow{3}{*}{Learnability}      & LY1       & 0.696    & 0.0331 & 21.0 & $<$.001 & 0.714           \\ \cline{2-7} 
                                   & LY2       & 0.733    & 0.0386 & 19.0 & $<$.001 & 0.664           \\ \cline{2-7} 
                                   & LY3       & 0.610    & 0.0403 & 15.1 & $<$.001 & 0.548           \\ \hline
\multirow{4}{*}{Operability}       & OP1       & 0.739    & 0.0300 & 24.7 & $<$.001 & 0.773           \\ \cline{2-7} 
                                   & OP2       & 0.728    & 0.0290 & 25.1 & $<$.001 & 0.782           \\ \cline{2-7} 
                                   & OP3       & 0.712    & 0.0326 & 21.9 & $<$.001 & 0.708           \\ \cline{2-7} 
                                   & OP4       & 0.620    & 0.0320 & 19.4 & $<$.001 & 0.645           \\ \hline
\multirow{4}{*}{Effectiveness}     & ES1       & 0.612    & 0.0352 & 17.4 & $<$.001 & 0.591           \\ \cline{2-7} 
                                   & ES2       & 0.741    & 0.0329 & 22.5 & $<$.001 & 0.721           \\ \cline{2-7} 
                                   & ES3       & 0.806    & 0.0339 & 23.8 & $<$.001 & 0.754           \\ \cline{2-7} 
                                   & ES4       & 0.776    & 0.0319 & 24.3 & $<$.001 & 0.761           \\ \hline
\multirow{4}{*}{Memorability}      & MY1       & 0.825    & 0.0311 & 26.5 & $<$.001 & 0.804           \\ \cline{2-7} 
                                   & MY2       & 0.857    & 0.0313 & 27.4 & $<$.001 & 0.822           \\ \cline{2-7} 
                                   & MY3       & 0.720    & 0.0310 & 23.2 & $<$.001 & 0.733           \\ \cline{2-7} 
                                   & MY4       & 0.677    & 0.0327 & 20.7 & $<$.001 & 0.673           \\ \hline
\multirow{5}{*}{Understandability} & UY1       & 0.715    & 0.0362 & 19.8 & $<$.001 & 0.645           \\ \cline{2-7} 
                                   & UY2       & 0.812    & 0.0311 & 26.1 & $<$.001 & 0.792           \\ \cline{2-7} 
                                   & UY3       & 0.782    & 0.0308 & 25.4 & $<$.001 & 0.776           \\ \cline{2-7} 
                                   & UY4       & 0.773    & 0.0327 & 23.6 & $<$.001 & 0.738           \\ \cline{2-7} 
                                   & UY5       & 0.747    & 0.0335 & 22.3 & $<$.001 & 0.709           \\ \hline
\end{tabular}
\end{table}

In addition, the Z-statistic (greater than 15) and p-value (less than 0.05) for each feature indicate statistical significance.

Next, we have investigated the model fit. In this regard, we first used the chi-squared statistic. The chi square-value of 1588 is large and highly significant (p$<$.001). However, the chi square-statistic used for assessing model fit is pretty sensitive to sample size, meaning that with a large sample a good enough fit between the model and the data almost always produces a large and significant (p$<$.05) chi square-value. 

Therefore, we need to examine other measures of model fit. Note that, the Comparative Fit Index (CFI) and the Tucker-Lewis Index (TLI) do not vary much with sample size. These two fit indices indicate a good fit, as the CFI of 0.924 and the TLI of 0.914 are both higher and exceed the 0.90 threshold. In addition, the Standardized Root Mean Square (SRMR) value of 0.0442 also suggests a good model fit as it is lower than the suggested $<$.08. In addition, the estimated value of the Root Mean Square Error of Approximation (RMSEA) of 0.0561 is close to 0, also indicating a good fit.

In summary, the results obtained from the analysis allow to confirm the construct validity of the underlying factor structures, extracted based on the prior research. 

In the light of the frequency, reliability, and validity findings discussed above, we find these results satisfactory and therefore proceed with further analysis to answer the stated research question: \textbf{What factors, and to what extent, directly influence the perceived usability of mobile applications?}

Table~\ref{tab:attributes-importance} shows eight usability factors, with a detailed breakdown, including the mean (\textit{M}), standard deviation (\textit{SD}), and coefficient of variation (\textit{CV}), along with the results of reliability ($\alpha$ and $\omega$) and convergent validity tests (\textit{AVE}).
Based on data collected from 838 respondents, efficiency exhibited the strongest influence, with a mean value of 4.07. The remaining attributes showed a moderate level of importance and were ranked in descending order as follows: Cognitive Load (3.73), Errors~(3.72), Learnability~(3.62), Operability~(3.60), Effectiveness~(3.58), Memorability~(3.46), and Understandability~(3.30).
Note that the coefficient of variation, ranging from 21.5\% to 25.2\%, suggests a moderate level of agreement among respondents.

\begin{table}[]
\small
\caption{The estimated values of the importance of the mobile usability factors, along with the results of reliability and convergent validity tests.}
\label{tab:attributes-importance}
\begin{tabular}{|l|l|l|l|l|l|l|}
\hline
\textbf{Factor}  & \textbf{M}    & \textbf{SD}  & \textbf{CV} & \textbf{$\alpha$} & \textbf{$\omega$} & \textbf{AVE} \\ \hline
Efficiency        & 4.07 & 0.951 & 0.234 & 0.916        & 0.917        & 0.734 \\ \hline
Cognitive Load    & 3.73 & 0.915 & 0.245 & 0.868        & 0.869        & 0.625 \\ \hline
Errors            & 3.72 & 0.869 & 0.234 & 0.850        & 0.851        & 0.586 \\ \hline
Learnability      & 3.62 & 0.832 & 0.230 & 0.679        & 0.687        & 0.417 \\ \hline
Operability       & 3.60  & 0.774 & 0.215 & 0.816        & 0.819        & 0.532 \\ \hline
Effectiveness     & 3.58 & 0.815 & 0.228 & 0.792        & 0.798        & 0.504 \\ \hline
Memorability      & 3.46 & 0.836 & 0.242 & 0.842        & 0.844        & 0.578 \\ \hline
Understandability & 3.30  & 0.830  & 0.252 & 0.851        & 0.853        & 0.539 \\ \hline
\end{tabular}
\end{table}

Considering the reliability statistics ($\alpha$ and $\omega$), we obtained values that met the minimum threshold of 0.6, implying acceptable reliability in the case of learnability, while good (between 0.70 and 0.89) or excellent (between 0.90 and 0.95) for the others. 

While the average variance extracted (AVE) should not be less than 0.5 to demonstrate an acceptable level of convergent validity, all factors except Learnability meet this criterion, meaning that the indicators (features) explain at least 50 percent of the variance in the underlying latent factor.
Nevertheless, despite certain weaknesses in indicator reliability and convergent validity, the measurement model is considered acceptable.

\section{Discussion}
\subsection{Implications for theory}
Our results have confirmed the importance of the eight attributes on the perceived usability of mobile applications derived from the PACMAD+3. Our results have confirmed the importance of the eight attributes for the perceived usability of mobile applications derived from PACMAD+3. However, their importance to our respondents differs from the general opinion where effectiveness, efficiency, and satisfaction are highlighted as the three most important factors \cite{johnson2020usability}. Obviously, this does not mean that existing mobile usability theory needs to be revisited. 

On the contrary, the importance of individual utility attributes may depend on many different determinants. During the data collection, the author personally collected many valuable verbal comments from the respondents. For example, for mobile gamers, the duration of launching an application is not that important, since by design a game launched on a mobile device takes relatively more time compared to an everyday mobile application such as an email client or a photo gallery. Thus, the generalizability of the results should be considered as the first limitation of our study. 

On the other hand, throughout our discussion we have both explicitly and implicitly emphasized the need to specify usability in a context of use. In addition, considering the limited functionality of mobile applications compared to their desktop equivalents, the variety of usability issues has unexpectedly magnified \cite{vasa2012preliminary, genc2017systematic, alqahtani2020insights} over time. For this reason, the ongoing discussion about mobile usability modeling is not a closed issue.  Indeed, considerable efforts have been made to address various usability issues \cite{heo2009framework, hensher2021scoping, schlichtig2024building}. Nevertheless, our study is based on a specific, newly introduced model, namely PACMAD+3.  While its generic nature makes it applicable in a wide scope, the contextless nature of the investigated attributes can be seen as a second limitation of our study. 

Evidently, more research is needed to fill these gaps. While other data samples may yield more evidence-based findings, the contextual settings may reveal hidden facets of the qualities that users expect from specific mobile solutions. It should be noted that the PACMAD+3 model does not impose any restrictions on research. On the contrary, it is an open reference model, without fixed relationships between attributes, nor limited by a fixed feature design. Thus, further research can cover the entire set of attributes or just a subset. In this line of thinking, a similar approach can be applied to feature selection.

\subsection{Implications for practice}
Our findings show that respondents praised all efficiency features the most. Actually, users expect fast performance, which means an efficient app responds quickly to input. On the other hand, an efficient app provides a seamless experience that keeps users active and engaged. Undeniably, applications that are fast and easy to use encourage frequent interaction, which can lead to higher user retention rates \cite{zuniga2019tortoise}. By focusing on efficiency, software vendors can develop mobile applications that are not only comfortable to use, but also reliable enough to reach a wider audience. But as many lessons learned show \cite{leitner2007usability, redlarski2016hard, inupakutika2022performance}, efficiency is not a one-time goal. On the contrary, it is an ongoing process in app development. Therefore, one should consider implementing continuous testing and optimization practices throughout the lifecycle.

On top of that, every task must be executable, regardless of the user's knowledge and skill level. Desirably, a task should require as few steps as possible to achieve the desired result. Therefore, the ultimate goal should be zero steps, which means eliminating or reducing user's workload by combining two or more tasks into one. That being said, effectiveness is measured by the percentage of tasks completed \cite{ferreira2022impact}, and more specifically by the number of steps required \cite{weichbroth2020usability}. In practice, effective user interface design follows three principles: utility, consistency, and accessibility \cite{UX-principles2023}. While in the first place usually workshops are organized with the participation of the actual users to get and better understand their needs and requirements \cite{canbazoglu2016developing}, then cognitive walkthroughs are used to identify usability issues from the perspective of user actions while using a mobile application \cite{jadhav2013usability}.

In this line of thinking, however in a broader view, the goal of the usability evaluation is to determine whether the user is likely to succeed at each step of the tasks performed, particularly with the goal of identifying and documenting any errors that occur. On the other hand, a cognitive walkthrough is a technique used to evaluate the learnability \cite{Salazar2022}. Since quality of use over time involves comparing the usability of a mobile application for experienced and novice users \cite{marrella2018measuring}, targeted testing is required when redesigning the user interface, or implementing additional features. For many software practitioners, the first imperative is simplified navigation, meaning easy and intuitive navigation, as a key driver for engaging and retaining users \cite{Shweta2024}.

While operability, memorability, and understandability are interrelated notions, mutually reinforcing to overall usability, interaction design appears to play a major role in the user's perception of its underlying logic and meaning. Pointing out the need for user testing for mobile applications is certainly not new.  However, interaction with mobile applications presents several challenges. First, while interaction with desktop computers relies on a mouse and keyboard, which provide more precise control but less direct engagement with screen objects, mobile users interact with touch gestures such as tapping, pinching, and swiping, which provide a more direct and tactile experience. Second, mobile (smaller) screens require more concise content and adaptive layouts that support both portrait and landscape orientation. Third, mobile applications rely on virtual keyboards and voice input, which can be less efficient for text-heavy tasks but more convenient for short inputs.

Note that from a user's perspective, the importance of certain qualities differs between application domains. Therefore, in usability evaluation settings, domain-specific features are typically prioritized, while others may be relaxed, or even omitted. For example, the response time of a mobile game in a multiplayer and multi-threaded scenario is usually not a concern, while the response time of an application that registers products during order picking is critical to the productivity of a warehouse worker. Similarly, while user satisfaction is paramount for the former, it is simply irrelevant for the latter.

To summarize. Any application that fails to satisfy users is considered a failure. Therefore, the development process should be approached with one goal in mind: maximum application usability and user satisfaction. The easiest way to establish a long-term strategy is to perform usability testing with clearly defined usability metrics. It seems that not only observable properties play an important role for the user, but also latent variables that are not always easy or unambiguous to measure. Therefore, in practice, the presented operationalization of the PACMAD+3 model should be adapted, taking into account the actual application features, as well as the current needs of the user. 

\subsection{Study Contributions}
Our contributions to the field of mobile human-computer interaction can be summarized in two distinct but mutually complementary parts. The first is the operationalization of the factors that constitute the PACMAD+3 mobile usability model. To the best of our knowledge, this is the first attempt to date that has ever been made. In addition, while typically latent variables are operationalized by meaningful statements, yet either intangible and untied to a specific application property or feature, our study stays in opposite to such approach, whenever the nature of a particular factor pretends its operationalization through the observable and tangible qualities or objects. 

The second contribution is an evaluation of mobile usability factors and corresponding features based on survey data collected from 838 respondents. Specifically, we provide evidence-based results showing their level of influence on the perceived usability of mobile applications. Such knowledge is valuable for at least two reasons. From a theoretical perspective, the strength and direction of the effects of individual features have been empirically demonstrated, while from a practical view, designers and developers can adopt and adapt specific features to enable users to perform tasks more efficiently.

\subsection{Threats to validity}
Since we used convenience sampling, not all members of the population have an equal opportunity to participate in the study. Therefore, our study has limited generalizability as the sample may not be representative of the broader population. In addition, convenience sampling often introduces selection bias, where certain demographic groups, user experience levels, or device preferences are overrepresented, leading to biased results. This can affect the reliability of usability assessments, as the sampled population may not reflect the diversity of real-world mobile users. 

However, to mitigate both of these threats, we collected data through multiple channels, including phone contacts and peers on social media networks. In this way, our respondents represented different professional groups, occupied different roles, and varied in gender, age, and education level. On the other hand, respondents reported using a variety of mobile applications in both their personal and professional spheres. 

The threat of invalid responses was mitigated by making the survey voluntary, with no incentives promised at the beginning of the survey. The majority of the questionnaires were submitted in paper form in the physical presence of the author. A preliminary screening was carried out under favorable conditions. In this way, a few cases were excluded from the sample for at least one of the following reasons: skipping one or more questions, selecting all available answers, or giving false answers, regarding age, work experience, or indicating the use of non-existent mobile applications. Furthermore, while it was technically possible to complete the survey more than once, we found no evidence of this in the data. Again, due to the effort required to complete the survey, we consider multiple entries from the same respondent to be highly unlikely.

The threat of bias in item wording was mitigated during the design process. The survey included 32 features expressed in natural language. Their names were taken from the relevant literature, including research and review papers, books, and gray literature such as reports, working papers, and websites. Finally, the coded themes were validated by referencing related studies that also used the same terms in a similar context. In addition, we also randomly asked respondents, immediately after returning the completed questionnaire, if any items were misunderstood.

The threat associated with the 5-point Likert scale used involves fewer response choices, resulting in a narrower range of opinions collected. However, the 5-point scale is said to be easy for people to understand and use \cite{arkorful2022voter} and is considered as one of the most widely used \cite{ginters2023pragmatic}.
In particular, a neutral response seems to be more tangible, ensuring that respondents don't lose interest. The Likert scale was also provided with a corresponding numerical descriptors. In addition, in order to reduce extreme response bias, neither positive nor negative emotional traits were contained in the survey items.

Another threat to validity is social desirability bias, whereby respondents may rate features as more important than they actually are in order to appear socially desirable. This risk was mitigated by ensuring voluntary and anonymous participation, as well as by not offering incentives, thereby reducing pressure to provide socially acceptable responses. Additionally, the survey items were written in neutral, non-evaluative language, and moral or normative cues implying “correct” answers were avoided. The presence of a clearly defined neutral midpoint on the five-point Likert scale allowed respondents to express indifference or uncertainty rather than forced agreement. Finally, the moderate coefficients of variation observed across attributes suggest meaningful variability in the collected responses, indicating that participants did not uniformly rate all features as highly important.

In general, while convenience sampling offers efficiency, its lack of representativeness may limit the external validity of findings. The use of stratified, quota, or random sampling can improve generalizability but still has inherent limitations. Therefore, other researchers may also attempt to use a hybrid approach, combining multiple sampling approaches depending on the research constraints, which may be the most effective strategy for achieving both practicality and robust findings.

\subsection{Future Research Directions}
As mobile applications have become an integral part of everyday life, they now fall under the category of places of public accommodation \cite{Visser2024}, making accessibility a relevant and timely consideration for future research. The most common accessibility obstacles for mobile applications include audio and video limitations \cite{ballantyne2018study}, low contrast and small text \cite{gil2020automatic}, inaccessible touch targets \cite{matos2023evaluation}, motion sensitivity \cite{Restack2025}, and poor screen reader support \cite{park2019development}. 
Besides, an active participation of people with auditory, cognitive, mobility, or visual impairments in usability testing requires a suitably adapted physical environment, hardware equipment and the necessary software capabilities.

With the advent of the 5th generation mobile network, which offers increased availability, ultra-low latency, higher capacity and increased bandwidth compared to 4G \cite{Intel2025}, the development of new applications and use cases for mobile devices has emerged. However, 5G has a profound impact on the mobile ecosystem \cite{ruiz2022mobile}, not only raising integration issues with other systems, including autonomous cars \cite{raddo2020end}, home automation devices \cite{gupta2017fog}, portable and wearable computing devices \cite{vermesan2022internet}, but also changing the user's view on the perceived usability of the mobile applications used. Considering a bunch of different devices and made by different manufacturers, controlled using one mobile solution opens new frontiers to explore with respect to user interface design patterns and interaction models. 

Undoubtedly, the next milestone in mobile human–computer interaction is the use of machine learning techniques for real-time user interface personalization based on both explicit and implicit features. One example is a news reader that delivers a customized stream of recommendations based on the user’s interests or location by ad hoc processing and aggregating data from multiple sources~\cite{constantinides2016user}. Other examples include emerging state-of-the-art AI-based technologies capable of recognizing voice commands, gestures, and sensor data. Once the appropriate permissions are granted, on-board AI agents can perform tasks on behalf of the user.

We have entered a new era, one in which AI agents, particularly generative AI (genAI), are powerful tools for content creation, as well as an era of the “Zero User Interface”. Unlike traditional interfaces with screens and buttons, zero-UI mobile applications rely on natural interaction modalities such as voice, context awareness, and individual preferences. Such applications can also collaborate with other intelligent agents by exchanging information to better respond to users’ needs.

Another two innovations that radically have changed the way mobile applications are used concern Augmented Reality (AR) and Virtual Reality (VR).  By design, AR enhances the user experience by using computer vision and object recognition to superimpose digital content, including audio, video, and graphics elements, onto users’ real environments \cite{sung2021effects}. VR goes further by providing a simulated experience by using 3D near-eye displays and poses tracking, giving the user an ubiquitous and high-fidelity experience of a virtual world \cite{tan2018supporting}. 

With the emergence of a new era of mobile applications and the continued blurring of the boundary between the real and digital worlds, the current concept of usability appears to require considerable revision. The proposed directions for future research are not exhaustive but instead reflect recent technological advancements.

\section{Conclusions}
Undeniably, usability in general stands for user acceptance of mobile applications \cite{alqahtani2015investigation, hajesmaeel2022most, camilleri2023functionality}. However, if one asks about the importance of certain factors that contribute to mobile usability, there are many voices around us that convincingly tell us which ones are relevant. 
Our study brings a better understanding of usability for mobile applications by evaluating the influence of eight factors borrowed from the PACMAD+3 model. In our opinion, the reported findings are useful for both researchers and practitioners, contributing to theory and practice by confirming the validity and importance of the model. It should be noted that our results also raise the possibility of using evaluated features in quality assurance strategy, at the top of the mobile software development process. Nevertheless, our paper does not pretend to be the last word on the subject, but further studies are needed to extend the frontiers of knowledge in this area. 

Undeniably, the perception of mobile usability has changed irrevocably with the growing adoption of AI-driven features. Recent releases of mobile applications goes beyond traditional tapping and swiping, incorporating voice interaction, computer vision, and adaptive responses. These advances are now pushing the boundaries of human-device interaction to new frontiers. There is no doubt that the next generation of mobile applications equipped with intelligent front-end features will require a new approach to usability testing, opening up an untold field of future research. 

While the transition to intelligent mobile applications undoubtedly brings significant benefits \cite{Christie2024}, it also raises concerns about user privacy and security \cite{elahi2021characterization}. It seems that the emerging revolution may revalue and reformulate the principles and patterns of mobile user interface design and development, ultimately and irrevocably changing the way we now understand usability today.

%\bibliographystyle{elsarticle-num}
%\bibliography{references}

\section*{Declarations}
\subsection*{Availability of data and materials }
Data available on reasonable request. Please contact correspondence author for access to dataset.
All data supporting the findings of this study are available within the paper.

\subsection*{Declaration of Interest Statement}
The authors declare that they have no known competing financial interests or personal relationships that could have appeared to influence the work reported in this paper.

\subsection*{Ethical Approval} 
No experiments were performed. Therefore, ethics committee approval was not required. Informed consent was obtained from all individual participants in the study. No experimental protocols were used. All research was performed in accordance with the guidelines and regulations of the Gdansk University of Technology. Written informed consent was obtained from all respondents for their participation in the study. All respondents were informed in writing that the collected data would be published. The data and test results in the manuscript cannot be linked to individual participants, as all questionnaires were anonymized.

\subsection*{Funding}
No funding has been received.

\subsection*{Author Contribution}
Pawel Weichbroth: Conceptualization; Methodology; Software; Validation; Formal analysis; Investigation; Resources; Data Curation; Writing - Original Draft; Writing - Review \& Editing; Visualization; Supervision; Project administration; Funding acquisition.

\clearpage
\end{document}